\documentclass[letter, twocolumn]{jpsj3}

\allowdisplaybreaks[1]

\usepackage{txfonts}
\usepackage{graphicx}
\usepackage{bm}
\usepackage{braket}
\usepackage{cases}
\usepackage{mathrsfs}
\usepackage{amsmath}
\usepackage{color}

\newcommand{\average}[1]{\ensuremath{\langle#1\rangle} }

\title{Generation of Spin Current from Lattice Distortion Dynamics: Spin-Orbit Routes}
\author{Takumi Funato and Hiroshi Kohno}
\inst{
 Department of Physics, Nagoya University, Nagoya 464-8602, Japan
}

\abst{
Generation of spin current from lattice distortion dynamics in metals 
is studied with special attention on the effect of spin-orbit coupling. 
 Treating the lattice distortion by local coordinate transformation, 
we calculate spin current and spin accumulation with the linear response theory. 
 It is found that there are two routes to the spin-current generation:  
one via the spin Hall effect and the other via the spin accumulation. 
 The present effect due to spin-orbit coupling can be comparable to, 
or even larger than, the one based on the spin-vorticity coupling 
in systems with strong spin-orbit coupling. 
}

\begin{document}

\maketitle

 In the field of spintronics, spin current occupies a central position for the development of new devices 
or the discovery of novel physical phenomena. 
 To date we know several methods available to generate spin currents, 
which include spin pumping\cite{pump01,pump02,pump1}, spin Hall effect\cite{SHE1}, 
spin accumulation at the ferromagnet/nonmagnet interface\cite{acc1},  
and spin Seebeck effect\cite{sse1}. 
 These are classified as magnetic, electrical, magnetoelectric, and thermal means, respectively.

 Recently, there has also been interest in generating spin currents by mechanical means, 
namely, by converting angular momentum associated with mechanical motion, 
such as the rigid rotation of a solid or vorticity of a fluid, 
into spin angular momentum of electrons.  
 In the experiments reported so far, two mechanisms have been considered. 
 One is the acoustic spin pumping by Uchida {\it et al.} \cite{asp1,asp2}, 
which is based on the magnon-phonon coupling. 
 They succeeded in generating spin current by injecting acoustic waves into yttrium iron garnet (YIG) 
from the attached piezoelectric element.
 Theoretical analyses were given by Adachi and Maekawa \cite{asptr1, aspt1}, 
Keshtgar {\it et al.} \cite{aspt2}, and Deymier {\it et al.} \cite{aspt3}.
 Another mechanism, proposed by Matsuo {\it et al.} \cite{srct1,srct2}, 
is based on the spin-rotation coupling or the spin-vorticity coupling (SVC). 
 This is the coupling of the spin to the effective magnetic field that emerges 
in a rotating (non-inertial) frame of reference locally fixed on the material that is in motion. 
The first experiment for the SVC mechanism was conducted on liquid metals.\cite{svce1,svct1} 
 To realize the SVC mechanism in solids,  
it was proposed to use surface acoustic waves \cite{sawt1,sawt2}. 
 Nozaki {\it et al.} used Py/Cu bilayer and injected surface acoustic waves 
into Cu from the attached $\text{LiNbO}_3$ (surface acoustic wave filter) \cite{nozaki}. 
 The generated AC spin current was detected via the spin-torque ferromagnetic resonance.

 One of the reasons that the mechanical generation of spin current has attracted attention is that 
it does not rely on spin-orbit interaction (SOI). 
 Therefore, previous works did not pay attention to the effects of SOI.  
 However, it is well expected that SOI plays certain roles in the mechanical processes 
of spin-current generation. 
 For example, the previous experiments\cite{asp1,asp2} were conducted on systems with an interface, 
which potentially possesses Rashba SOI. 
 Furthermore, the mechanical generation method may be used in combination with other 
\lq\lq conventional'' mechanisms that utilize SOI, and thereby enhance spin current.

 In this paper, we study a mechanical generation of spin current by focusing on the effects of SOI.  
 As a mechanical process, we consider dynamical lattice deformations of a solid with metallic electrons and with SOI. 
 To treat lattice deformations analytically, we use the method of Tsuneto developed in the context of ultrasonic attenuation in superconductors \cite{tsuneto1}, 
which employs a local coordinate transformation. 
 By calculating spin current and spin accumulation induced by dynamical lattice deformations, 
we found two routes to spin-current generation:  
one via the spin accumulation and the other via the spin Hall effect. 
 As a related work, Wang {\it et al}. \cite{soi1} derived the Hamiltonian that includes SOI 
in a general coordinate system starting from the general relativistic Dirac equation, 
but they did not give an explicit analysis of spin-current generation.

{\it Model:} \
 We consider a free-electron system in the presence of random impurities and the associated SOI. 
 The Hamiltonian is given by 
\begin{align}
H = -\frac{\nabla '^2}{2m} + V_{\text{imp}}({\bm r}') 
     + i\lambda_{\text{so}} \{ [\nabla ' V_{\text{imp}}({\bm r}')] \times {\bm \sigma } \} \cdot \nabla ' . 
\label{lab.H}
\end{align}
 The second term represents the impurity potential, 
$V_{\text{imp}}({\bm r}') = u_{\rm i} \sum_j \delta ({\bm r}'-{\bm R}_j)$, 
with strength $u_{\rm i} $ and at position ${\bm R}_j$ (for $j$th impurity), 
and the third term is the SOI associated with $V_{\text{imp}}$, 
with strength $\lambda _{\text{so}}$ and the Pauli matrices 
${\bm \sigma }=(\sigma ^x, \sigma ^y, \sigma ^z)$. 
 When the lattice is deformed, e.g., by sound waves, the Hamiltonian becomes 
\begin{align}
 H_{\text{lab}} = &-\frac{\nabla '^2}{2m} + V_{\text{imp}}({\bm r}'-\delta {\bm R}({\bm r}',t)) 
\nonumber
\\
&+ i\lambda _{\text{so}} \left\{ \Bigl[ \nabla ' V_{\text{imp}}\left( {\bm r}'-\delta {\bm R}({\bm r}',t)\right) \Bigr] \times {\bm \sigma }\right\} \cdot \nabla ',
\end{align}
where $\delta {\bm R}({\bm r}', t)$ is the displacement vector of the lattice 
from their equilibrium position ${\bm r}'$.

 Following Tsuneto \cite{tsuneto1}, we make a local coordinate transformation, 
${\bm r}={\bm r}'-\delta {\bm R}({\bm r}', t)$, from the laboratory (Lab) frame (with coordinate ${\bm r}'$) 
to a \lq\lq material frame" (with coordinate ${\bm r}$) which is fixed to the \lq atoms' 
in a deformable lattice. 
 At the same time, the wave function needs to be redefined to keep the normalization condition,  
\begin{align}
 \psi ({\bm r}, t)=[1+\nabla \cdot \delta {\bm R}]^{1/2}\psi '({\bm r}',t) + \mathscr{O}(\delta R ^2),
\end{align}
where $\psi ' ({\bm r} ', t)$ is the wave function in the Lab frame, and $\psi ({\bm r}, t)$ is the one in the material frame. 
Up to the first order in $\delta {\bm R}$, the Hamiltonian for $\psi ({\bm r}, t)$ is given by 
\begin{align}
H_{\text{mat}} = H + H'_{\text{K}} + H'_{\text{so}},
\end{align} 
where $ H = H_{\text K} + H_{\text{imp}} + H_{\text{so}} $
is the unperturbed Hamiltonian defined by $H_{\text{lab}}$ with $\delta {\bm R}={\bm 0}$. 
 Here, $H_{\rm K} = \sum_{\bm k} (k^2/2m) \psi ^{\dagger}_{\bm k} \psi _{\bm k}$ is the kinetic energy, 
with $\psi _{\bm k}$ ($\psi ^{\dagger}_{\bm k}$) being the electron annihilation (creation) operator. 
$H_{\text{imp}}$ and $H_{\text{so}}$ describe the impurity potential and impurity SOI, respectively, 
$ H_{\text{imp}} =\sum_{\bm k , \bm k'}V_{{\bm k}'-{\bm k}} \psi ^{\dagger}_{{\bm k}'} \psi _{\bm k}$,
$ H_{\text{so}} = i\lambda _{\text{so}} \sum_{\bm k , \bm k'}V_{{\bm k}'-{\bm k}} 
 ({\bm k}'\times {\bm k}) \cdot \psi ^{\dagger}_{\bm k'}{\bm \sigma} \psi _{\bm k}$,
where $V_{{\bm k}'-{\bm k}}$ is the Fourier component of $V_{\text{imp}}({\bm r})$.
 Assuming a uniformly random distribution, we average over the impurity positions as 
$\average{V_{\bm k}V_{\bm k'}}_{\text{av}} = n_{\text i} u_{\rm i} ^2 \delta _{{\bm k}+{\bm k}', {\bm 0}}$, 
and 
$\average{V_{\bm k}V_{\bm k'}V_{\bm k''}}_{\text{av}} 
 = n_{\text i} u_{\rm i} ^3 \delta _{\bm k+\bm k'+\bm k'',0}$,
where $n_{\text i}$ is the impurity concentration.
 The impurity-averaged retarded/advanced Green function is given by
$G_{\bm k}^{\rm R/A} ( \varepsilon ) = (\varepsilon + \mu - k^2/2m \pm  i\gamma )^{-1}$,
where 
$\gamma = \pi n_{\text i} u_{\rm i}^2N(\mu ) (1+ \frac{2}{3} \lambda _{\text{so}}^2k_{\text F}^4)$ 
is the damping rate. 
 Here, $N(\mu)$ is the Fermi-level density of states (per spin), 
and $k_{\rm F}$ is the Fermi wave number. 
 In this work, we consider the effects of SOI up to the second order.

\begin{figure}[tbp]
  \begin{center}
   \includegraphics[width=55mm]{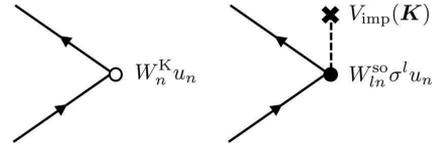}
  \end{center}
 \caption{Two types of vertices associated with the coupling to the lattice displacement $\delta {\bm R}$, 
 or the velocity field ${\bm u} = d\delta {\bm R}/dt$.}
 \label{vertices}
\end{figure}

The effects of lattice distortion are contained in $H'_{\text{K}}$ and $H'_{\text{so}}$, which come from $H_{\rm K}$ and $H_{\rm so}$, respectively.
In the first order in $\delta {\bm R}$, they are given by
\begin{align}
 H'_{\text K} 
&= \sum_{\bm k} W^{\text K}_{n}({\bm k})  \, u^n_{\bm q, \omega} \,  
    \psi _{\bm k+\frac{\bm q}{2}}^{\dagger}\psi_{\bm k-\frac{\bm q}{2}}  ,
\\
 H'_{\text{so}} 
&= \sum _{\bm k , \bm k'}  
  V_{{\bm k}'-{\bm k}} W^{\text{so}}_{ln}({\bm k},{\bm k}') \, u^n_{\bm q, \omega} \,  
  \psi ^{\dagger}_{{\bm k}'+\frac{\bm q}{2}} \sigma ^l \psi _{{\bm k}-\frac{\bm q}{2}} . 
\end{align}
 Here, ${\bm u}_{\bm q, \omega}$ is the Fourier component of the lattice velocity field, 
${\bm u}({\bm r}, t) = \partial _t \delta {\bm R} ({\bm r}, t)$, 
and we defined (see Fig.~\ref{vertices}), 
\begin{align}
 W^{\text K}_{n}({\bm k}) &= \Bigl[ \frac{{\bm q}\cdot {\bm k}}{m\omega} - 1 \Bigr] \, k_{n}  , 
\label{vcsk}
\\
 W^{\text{so}}_{ln}({\bm k},{\bm k}') 
&= \frac{\lambda _{\text{so}} }{i\omega}    
 \Bigl[ ({\bm k}\times {\bm q}) _l \, k'_n - ({\bm k}'\times {\bm q}) _l \, k_n \Bigr] . 
 \label{vcso1}
\end{align}
 The first term in $W_n^{\rm K}$ describes the coupling of the strain $\partial_i \delta R_n$ 
to the stress tensor $\sim \!\! \sum_{\bm k} k_i k_n c_{\bm k}^\dagger c_{\bm k}$ of electrons, 
and modifies the effective mass tensor. 
 Throughout this report, ${\bm q}$ represents the wave vector of the lattice deformation 
and $\omega$ is its frequency. 
 We assume that the spatial and temporal variations of $\delta {\bm R}$ are slow and satisfy 
the conditions $q \ll \ell^{-1}$ and $\omega \ll \gamma$, where $\ell$ is the mean free path.

 Spin and spin-current density operators are given by
\begin{align}
 \hat{j}^\alpha_{\text s, 0}({\bm q}) 
&=\hat{\sigma} ^\alpha ({\bm q})  
 = \sum _{\bm k} \psi ^{\dagger}_{\bm k-\frac{\bm q}{2}} \sigma^\alpha v_0 \psi _{\bm k+\frac{\bm q}{2}},
\\
 \hat{j}^\alpha_{\text s, i}({\bm q}) 
&= \sum _{\bm k}  \psi ^{\dagger}_{\bm k-\frac{\bm q}{2}} \sigma^\alpha v_i \psi _{\bm k+\frac{\bm q}{2}} 
+ \hat{j}^{\text a, \alpha}_{\text s, i}({\bm q}),
\end{align}
where $\alpha = x,y,z$ specifies the spin direction, $i=x,y,z$ the current direction, and $v_0=1$.
Here,  
\begin{align}
 \hat{j}^{\text a, \alpha}_{\text s, i}({\bm q}) 
= -i \lambda_{\text{so}} \epsilon _{\alpha ij}  \sum_{{\bm k},{\bm k'}} V_{{\bm k}'-{\bm k}}  (k_j' -  k_j) \, 
 \psi^{\dagger}_{{\bm k}'-\frac{{\bm q}}{2}} \psi _{{\bm k} + \frac{{\bm q}}{2}},
 \label{ascop}
\end{align}
is the \lq anomalous' part of the spin-current density, with $\epsilon _{\alpha ij}$ being the Levi-Civita symbol.
We calculate $\hat{j}^\alpha_{\text s, \mu}$ in (linear) response \cite{kubo1} to ${\bm u}$, 
\begin{align}
 \average{\hat{j}^\alpha_{\text s, \mu}({\bm q})}_{\omega }
= -\Bigl[ K^{\text{ss}, \alpha}_{\mu n} + K^{\text{sj}, \alpha}_{\mu n} 
             +  K^{\text{so}, \alpha}_{\mu n} 
   \Bigr] _{{\bm q}, \omega} u_n , 
\end{align}
where $K^{\text{ss}, \alpha}_{\mu n}$ ($K^{\text{sj}, \alpha}_{\mu n}$) is the skew-scattering (side-jump) type 
contribution in response to $H'_{\text K}$, 
and $K^{\text{so}, \alpha}_{\mu n}$ describes the response to $H'_{\text{so}}$.

\begin{figure}[tbp]
  \begin{center}
   \includegraphics[width=85mm]{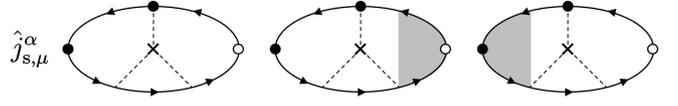}
  \end{center}
 \caption{Skew-scattering type contributions to the spin current ($\mu = x, y, z$) and/or 
             spin accumulation ($\mu =0$).
The black (white) circles represent spin-flip (spin non-flip) vertices.
 The cross and the dashed line represent an impurity and the impurity potential, respectively. 
 The shaded part represents the impurity ladder vertex corrections.
 The upside-down diagrams are also considered in the calculation.
}
 \label{skew}
\end{figure}

{\it Skew-scattering process:} \
 The skew-scattering contribution without ladder vertex corrections, shown in Fig.~\ref{skew}, is given by 
\begin{align}
 K^{\text{ss}, \alpha}_{\mu \nu}(\omega ) 
&= i\lambda _{\text{so}}n_{\text i} u_{\rm i}^3 \sum_{{\bm k}_1, {\bm k}_2}({\bm k}_1\times {\bm k}_2)^\alpha 
     v_{1\mu} W^{\text K}_{\nu}({\bm k}_2) 
\nonumber \\
&\times \frac{\omega}{i\pi} \sum_{\bm p} \left[ G^{\text R}_{\bm p}\left( \frac{\omega}{2}\right) - G^{\text A}_{\bm p}\left( -\frac{\omega}{2}\right) \right] 
\nonumber \\
&\times G^{\text R}_{{\bm k}_{1+}} G^{\text A}_{\bm k_{1-}} G^{\text R}_{{\bm k}_{2+}} G^{\text A}_{\bm k_{2-}} , 
\end{align}
with $G^{\text{R/A}}_{\bm k\pm}=G^{\text{R/A}}_{{\bm k}\pm \frac{\bm q}{2}}(\pm \frac{\omega}{2})$.
 By including ladder vertex corrections, spin accumulation and spin-current density are calculated 
as\cite{suppl}
\begin{align}
 \average{\sigma ^\alpha}^{\text{ss}} 
&=\alpha _{\text{SH}}^{\text{ss}} n_{\text e}\tau \left( \frac{3}{5}Dq^2-i\omega  \right) 
    \frac{(i{\bm q}\times {\bm u})^\alpha}{Dq^2-i\omega + \tau _{\text{sf}}^{-1}},
\label{ssspin}
\\
 \average{{ j}^\alpha_{\text s, i}}  ^{\text{ss}}
&=  -Diq_i \average{\sigma ^\alpha}^{\text{ss}} + \alpha _{\text{SH}}^{\text{ss}}   \epsilon _{\alpha im} \average{ j_m },
\label{sscurrent}
\end{align}
where $\alpha _{\text{SH}}^{\text{ss}}=\frac{2\pi}{3}k_{\text F}^2\lambda _{\text{so}}N(\mu ) u_{\rm i} $ 
is the spin Hall angle due to skew scattering \cite{com1}, 
$n_{\text e}=\frac{2}{3m} k_{\text F}^2N(\mu )$ is the electron number density, 
$\tau =(2\gamma)^{-1}$ is the scattering time, 
 $D=\frac{1}{3}v_{\text F}^2\tau $ is the diffusion constant, 
and $\tau _{\text{sf}}^{-1} = (4\lambda _{\text{so}}^2k_{\text F}^4/3) \, \tau^{-1}$ is the spin relaxation rate 
due to SOI. 
 In Eq.~(\ref{sscurrent}), $\average{ j_m }$ is the charge current, 
\begin{align}
 \average{ j_m }
&= n_{\text e} \tau \biggl\{ - \left( \frac{3}{5}Dq^2-i\omega  \right) u_m 
\nonumber \\
& \ \ \  + \left( \frac{6}{5} 
+ \frac{ 1}{\tau (Dq^2-i\omega )} \right) D iq_m ( i{\bm q} \cdot {\bm u} ) \biggr\} ,
\label{cc}
\end{align}
generated by ${\bm u}$\cite{tsuneto1}. 
 Here, in the first line, the term $\sim \!\! Dq^2 u_m$ (the term $\sim \! i\omega u_m$) 
is induced by the first (second) 
term in $W_n^{\rm K}$, Eq. (\ref{vcsk}), via the spatio-temporal variation of the strain tensor 
$\partial_i \delta R_m$ (temporal variation of the velocity field $u_m$).   
 The last term is the diffusion current. 
 We see that a spin accumulation (\ref{ssspin}) is induced by the vorticity of the lattice velocity field ${\bm u}$. 
 The first term and the second terms in Eq.~(\ref{sscurrent}) are written with the spin accumulation  (Eq.~(\ref{ssspin})) and the charge current (Eq.~(\ref{cc})), respectively.

\begin{figure}[tbp]
  \begin{center} 
   \includegraphics[width=85mm]{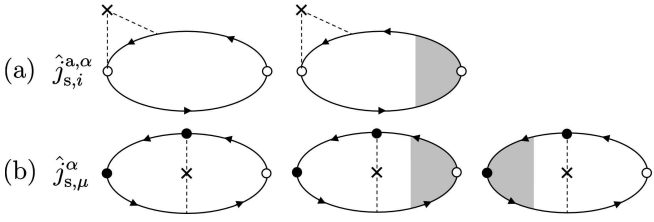} 
  \end{center}
 \caption{Side-jump type contribution to the spin-current density ($i, \mu = x, y, z$) 
  and/or spin accumulation $(\mu = 0)$. 
  The diagrams in (a) come from the anomalous velocity, Eq.~(\ref{ascop});  
  hence they contribute only to the spin current. 
 The diagrams in (b) can be nonvanishing only when the lattice deformation is nonuniform. 
 The upside-down diagrams are also included in the calculation.}
 \label{sj}
\end{figure}

{\it Side-jump process:} \
 The side-jump contributions are obtained from the two types of diagrams in Fig.~\ref{sj}. 
 They give
\begin{align}
 K^{\text{sj (a)}, \alpha}_{in}(\omega ) 
&= i\lambda _{\text{so}} n_{\text i} u_{\rm i} ^2 \epsilon_{\alpha ij} \sum_{\bm k_1, \bm k_1', \bm k_2} 
        (k_{1,j}'-k_{1,j}) W^{\text K}_n (\bm k_2)  
\nonumber \\
& \times \frac{\omega}{i\pi} 
  \Bigl[ \delta _{\bm k_1'\bm k_2}G_{{\bm k}_{1+}}^R + \delta _{\bm k_1\bm k_2} G_{{\bm k}_{1'-}}^A \Bigr] 
  G_{{\bm k}_{2+}}^R G_{{\bm k}_{2-}}^A  ,
  \\
   K^{\text{sj (b)}, \alpha}_{\mu \nu} (\omega ) 
&=  \lambda _{\text{so}} n_{\text i} u_{\rm i} ^2 
   \sum_{{\bm k}_1, {\bm k}_2}[({\bm k}_1- {\bm k}_2) \times  i{\bm q}]^\alpha  
   v_{1\mu} W^{\text K}_{\nu}(\bm k_2)  
\nonumber \\
&  \times \frac{\omega}{i\pi} 
   G_{{\bm k}_{1+}}^R G_{{\bm k}_{1-}}^AG_{{\bm k}_{2+}}^R G_{{\bm k}_{2-}}^A ,
\end{align}
corresponding to the diagrams in Fig.~\ref{sj} (a) and (b), respectively.
 With the ladder vertex corrections included, spin accumulation and spin-current density are calculated 
as\cite{suppl} 
\begin{align}
\average{\sigma ^\alpha}^{\text{sj}} 
&= \alpha ^{\text{sj}}_{\text{SH}} n_{\text e}\tau \left( \frac{3}{5}Dq^2 - i\omega \right) 
    \frac{(i{\bm q}\times {\bm u})^\alpha}{Dq^2 -i\omega + \tau _{\text{sf}}^{-1}}, 
\label{sdsj}
\\
 \average{j^\alpha_{\text s , i}}^{\text{sj (a)}} 
&= \alpha ^{\text{sj}}_{\text{SH}} \epsilon _{\alpha im} \average{j_m},
\label{scsj1}
\\
 \average{j^\alpha_{\text s, i}}^{\text{sj (b)}} 
&= \alpha ^{\text{sj}}_{\text{SH}} n_{\text e} \tau 
    \frac{\epsilon _{\alpha im} Diq_m}{Dq^2-i\omega} i{\bm q} \cdot {\bm u} 
    - Diq_i \average{\sigma ^\alpha}^{\text{sj}}.
 \label{scsj2}
\end{align}
where $\alpha^{\text{sj}}_{\text{SH}} = - \lambda_{\text{so}} m/ \tau$ is the spin Hall angle 
due to side-jump processes \cite{com1}.
 The diagrams in Fig.~\ref{sj} (a) give only the spin current (Eq.~(\ref{scsj1}))  
since the left vertices come from the anomalous velocity. 
 This contribution is also written with the charge current $\average{j_m}$ given by Eq.~(\ref{cc}).
 On the other hand, spin accumulation coming from Fig. \ref{sj} (b) is again proportional to the vorticity 
of the velocity field ${\bm u}$. 
 In Eq.~(\ref{scsj2}), the first term is proportional to the diffusion part of the charge current [the last term 
in Eq~(\ref{cc})], 
and the second term is the diffusion spin current. 
 We note that these contributions, coming from the diagrams of Fig.~\ref{sj} (b), vanish 
when the external perturbation is uniform, i.e., ${\bm q} = {\bm 0}$.

\begin{figure}[tbp]
  \begin{center}
   \includegraphics[width=60mm]{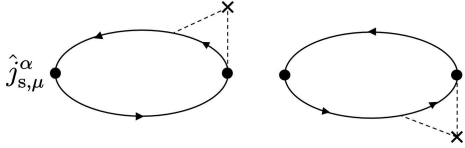}
  \end{center}
 \caption{Response to $H'_{\text{so}}$, which turned out to vanish.}
 \label{so}
\end{figure}

 Finally, the response to $H'_{\text{so}}$, shown in Fig.~\ref{so}, turned out to vanish, 
$K^{\text{so}, \alpha}_{\mu n} = 0$. 
 This is also the case when the ladder vertex corrections are included.

\begin{figure}[tbp]
  \begin{center}
   \includegraphics[width=85mm]{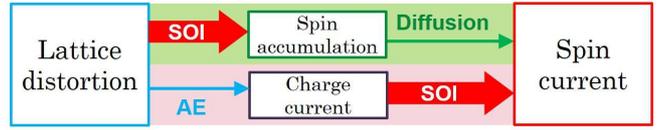}
  \end{center}
 \caption{(Color online) 
        Two routes to the generation of spin current from lattice distortion dynamics. 
        The thick arrows indicate the processes governed by SOI.
\lq AE' means acousto-electric effect.}
 \label{flow chart}
\end{figure}

{\it Result:} \
 Taken together, the total spin accumulation and spin-current density arising from the dynamical lattice distortion via SOI have been obtained as
\begin{align}
 \average{\sigma^\alpha}_{\text{SOI}} 
&= \alpha _{\text{SH}} n_{\text e}\tau \left( \frac{3}{5}Dq^2 - i\omega \right) \frac{(i{\bm q}\times {\bm u})^\alpha}{Dq^2 -i\omega + \tau _{\text{sf}}^{-1}}, 
\label{spin} 
\\
 \average{j^\alpha_{\text s, i}}_{\text{SOI}} 
&= - Diq_i \average{\sigma^\alpha}_{\text{SOI}} + \alpha _{\text{SH}} \epsilon _{\alpha im} \average{j_m} 
\nonumber \\
&+ \alpha ^{\text{sj}}_{\text{SH}} n_{\text e} \tau 
    \frac{\epsilon_{\alpha im} Diq_m}{Dq^2-i\omega} i{\bm q} \cdot {\bm u},
\label{sc}
\end{align}
where  $\alpha _{\text{SH}}=\alpha ^{\text{ss}}_{\text{SH}}+\alpha ^{\text{sj}}_{\text{SH}}$ 
is the \lq total' spin Hall angle.
 As seen from Eq.~(\ref{spin}), spin accumulation is induced by the vorticity of the lattice velocity field via SOI.
 The resulting diffusion spin current contributes to Eq.~(\ref{sc}) as the first term. 
 In addition, dynamical lattice distortion generates a charge current as well 
(known as the acousto-electric effect \cite{ae1}), which is then converted to a spin Hall current (in the transverse direction) via SOI, as expressed by the second and third terms in Eq.~(\ref{sc}).
 Therefore, there are two routes to the spin-current generation in the present mechanism;  
one is the \lq\lq diffusion route'' caused by the spin accumulation and the other is the \lq\lq spin Hall route'' that follows the acousto-electric effect.\cite{tatara1} 
 This is illustrated in Fig.~\ref{flow chart}. 
 In the latter (spin Hall) route, the longitudinal component of ${\bm u}$ also induces spin current 
via the generation of charge current.
 Finally, we note that the induced spin accumulation (\ref{spin}) and the spin-current density (\ref{sc})  
satisfy the spin continuity equation, 
\begin{align}
  \partial _t \average{\sigma^\alpha}_{\text{SOI}}  + \nabla \cdot \average{\bm j ^\alpha_{\text s}}_{\text{SOI}} 
= -\frac{\average{\sigma ^{\alpha}}_{\text{SOI}}}{\tau _{\text{sf}}} .
\end{align} 
 The term on the right-hand side represents spin relaxation due to SOI.

 The above result does not include the effects of lattice distortion on the spinorial character 
of the electron wave function. 
 Such effects are derived from the spin connection in the general relativistic Dirac equation.\cite{srct1} 
 The total spin current and spin accumulation are given by the sum of the contributions from the SVC 
(previous work\cite{srct1}) and SOI (present work). 
 Next, we study the contribution from SVC, an effect originating from the spin connection.

{\it Spin-rotation coupling:} \
 For comparing the present result with the previous one that is based on the spin-vorticity coupling 
(SVC) \cite{svct1}, 
we also calculate the spin accumulation and spin-current density in response to the vorticity 
of the lattice velocity field, 
${\bm \omega} = \nabla \times {\bm u}$.\cite{com_omega} 
 By treating the SVC Hamiltonian  
$H_{\text{SV}} = -\frac{1}{4}{\bm \sigma} \cdot {\bm \omega}({\bm q}, \omega )$  
as a perturbation, one has  
$  \average{j^\alpha_{\text s, \mu}}_{\omega} 
= \chi ^{\alpha}_{\mu \beta}({\bm q}, \omega ) \, \omega ^{\beta}({\bm q}, \omega )$, 
where $\chi ^{\alpha}_{\mu \beta}({\bm q}, \omega )$ is the response function.
 The response function 
(without vertex corrections) is given as
$\chi ^{\alpha}_{\mu \beta}({\bm q}, \omega ) 
 = \frac{1}{2}N(\mu ) \delta _{\alpha \beta} \delta _{\mu 0}   + \frac{i\omega}{4\pi} \delta _{\alpha \beta} \sum_{\bm k} v_\mu G^{\text R}_{\bm k_+}G^{\text A}_{\bm k_-}$.
 With ladder vertex corrections, spin accumulation and spin-current density are obtained as 
\begin{align}
 \average{\sigma^\alpha}_{\text{SV}} 
&= \frac{N(\mu )}{2} \frac{Dq^2 +\tau _{\text{sf}}^{-1}}{Dq^2-i\omega +\tau _{\text{sf}}^{-1}}  \omega^\alpha ,
\label{svspin}
\\
 \average{ j^\alpha_{\text s, i}}_{\text{SV}} 
&= -i\omega \frac{N(\mu )}{2} \frac{Diq_i}{Dq^2-i\omega + \tau _{\text{sf}}^{-1}} \omega^\alpha .
\label{svsc}
\end{align}
 They satisfy the spin continuity equation,  
\begin{align}
 (\partial _t + \tau _{\text{sf}}^{-1} ) \average{\sigma^\alpha}_{\text{SV}} + \nabla \cdot \average{{\bm j}^\alpha_{\text s}}_{\text{SV}} 
&= \frac{N(\mu )}{2 \tau _{\text{sf}}} \omega^\alpha , 
\label{svccon}
\end{align}
with a source term ($\, \sim \!\!\! {\bm \omega}$) on the right-hand side. 
 Alternatively, one may define 
the \lq\lq spin accumulation'' $\delta \mu^\alpha = \mu_\uparrow - \mu_\downarrow$ 
by $\average{\sigma^\alpha}_{\text{SV}} = n_\uparrow - n_\downarrow 
      = N(\mu) (\delta \mu^\alpha + \hbar \omega^\alpha /2)$,\cite{sawt1}  
where the spin quantization axis has been taken along the $\hat \alpha$ axis. 
 Then, Eq.~(\ref{svspin}) leads to 
\begin{align}
 ( \partial_t - D \nabla^2 + \tau _{\text{sf}}^{-1} ) \, \delta \mu^\alpha 
&= - \frac{\hbar}{2} \dot \omega^\alpha . 
\label{del_mu}
\end{align}
 This is the basic equation used in Ref.~16 to study spin-current generation. 
 Therefore, in the SVC mechanism, only the transverse acoustic waves generate spin current, 
and the generated spin current is purely of diffusion origin. 
 These are in stark contrast with the SOI-induced mechanism.

 {\it Comparison:} \
 To see the magnitude of the present effect,  
we estimate the diffusion spin current generated via SOI, Eq.~(\ref{spin}), relative to the one due to SVC, Eq.~(\ref{svspin}), 
\begin{align}
 R_{\text{diff}}(f)\equiv \left| \frac{\average{j^\alpha _{\text s, i}}_{\text{SOI}}^{\text{diff}}}{\average{j^\alpha_{\text s, i}}_{\text{SV}}} \right| 
= \frac{8}{3} \alpha _{\text{SH}} \frac{\varepsilon_{\text F} \tau}{\hbar} 
   \left| 1 + \frac{6 \pi i}{5} \frac{Df}{v_{\text a}^2} \right| , 
\label{compare}
\end{align} 
where $f=\omega /2\pi$ is the frequency and $v_{\text a} = \omega /q$ is the (phase) velocity 
of acoustic waves, $\varepsilon_{\text F} = \hbar^2 k_{\text F}^2/2m$ is the Fermi energy, 
and $\hbar$ has been recovered. 
 This ratio is larger for higher frequency $f$, and for materials with stronger SOI.

 For CuIr, the spin Hall angle is $2 \alpha _{\text{SH}}=2.1\pm 0.6\%$, independent of impurity 
concentration, 
which is dominated 
by the extrinsic, skew-scattering process\cite{niimi1}. 
 In the nearly free electron approximation with 
the Fermi wave number $k_{\text F}= 1.36 \times 10^{10}$ m$^{-1}$, 
Fermi velocity $v_{\text F}=1.57 \times 10^6$m/s, 
effective mass $m^* = 8.66 \times 10^{-31}$ kg \cite{Cudata}, 
and resistivity $\rho _{\text{imp}} = 7.5$ $\mu \Omega \text{cm}$ (for $3\%$ Ir),
we estimate the scattering time as $\tau _{\text{imp}}=5.30 \times 10^{-15}$ s,  
and the diffusion constant as $D_{\text{imp}}=4.35 \times 10^{-3}$ m$^2$/s, due to impurities.
 With the speed of the Rayleigh type surface acoustic wave,
$v_{\text a} = 3.80 \times 10^3$ m/s, on a single crystal of LiNbO$_3$\cite{danseihadata}, we obtain 
\begin{align}
 R_{\text{diff}}^{\text{CuIr}}(f) = 1.51 \sqrt{1 + (1.14 \times f )^2} , 
\end{align}
where $f$ is expressed in GHz.
 Therefore the diffusion spin current $\average{j^\alpha _{\text s, i}}_{\text{SOI}}^{\text{diff}}$ via SOI  
is comparable to, or even larger than, that from SVC in metals with strong SOI. 
 It is thus expected that the total contribution $\average{j^\alpha _{\text s, i}}_{\text{SOI}}$, 
which includes both the diffusion spin current and the spin Hall current, can be larger than 
$\average{j^\alpha_{\text s, i}}_{\text{SV}}$. 
 The magnitude itself is, however, small;  
$\langle j_{{\rm s}, z}^x \rangle = 10^{20} \sim 10^{24}$ m$^{-2}$s$^{-1}$ 
$= 10 \sim 10^5$A/m$^2$ 
for $\tau_{\rm sf}^{-1} = 0 \sim 5 \times 10^{13}$s$^{-1}$, $\delta R =1$\AA, and $f=3.8$ GHz, 
as in the case of the SVC mechanism\cite{srct2}.

 To summarize, we studied the generation of spin current and spin accumulation by dynamical 
lattice distortion in metals with SOI at the impurities. 
 We identified two routes to the spin-current generation, namely, 
 the \lq\lq spin Hall route'' and the \lq\lq spin diffusion route.''  
 In the former route, a charge current is first induced by dynamical lattice distortion, 
which is then converted into a spin Hall current.  
 In the latter route, a spin accumulation is first induced from the vorticity of the lattice velocity field, 
which then induces a diffusion spin current. 
 The result suggests that the spin accumulation (hence the associated diffusion spin current) 
generated via SOI is larger than that due to SVC for systems with strong SOI. 
 Similar effects are expected in systems with other types of SOI, such as Rashba, Weyl, etc., 
and such studies will be reported elsewhere. 
 In this connection, we note that Xu {\it et al.} recently reported an experiment on the mechanical 
spin-current generation (due to magnon-phonon coupling) in a system with Rashba SOI\cite{mingran}.

\acknowledgment 
 We would like to thank K. Kondou, J. Puebla, and M. Xu for the valuable and informative discussion, 
and J.~Ieda, S.~Maekawa, M.~Matsuo, M.~Mori, and K.~Yamamoto for their valuable advice and comments. 
 We also thank A. Yamakage, K. Nakazawa, T. Yamaguchi, Y. Imai, and J. Nakane for the daily discussions. 
 This work is supported by JSPS KAKENHI Grant Numbers 25400339, 15H05702 and 17H02929. 
 TF is supported by a Program for Leading Graduate Schools: 
\lq\lq Integrative Graduate Education and Research in Green Natural Sciences.''


\begin{thebibliography}{99}
\bibitem{com1} 
 We define the spin Hall angle as half of that of Ref.~\cite{SHE1} in order to be consistent 
with the definition of spin-current operators.
\bibitem{com_omega} While the same Greek letter is used for the external frequency ($\omega$) 
and the vorticity (${\bm \omega}$ or $\omega^\alpha$), we hope no confusion will arise 
since the latter is denoted as a vector.
\end{thebibliography}
\end{document}